\pdfoutput=1
\documentclass{biophys-new}
\usepackage[utf8]{inputenc}
\usepackage[colorlinks,allcolors=cyan!70!black]{hyperref}
\usepackage{amsmath}
\usepackage{graphicx}

\hypersetup{pdftitle={Quantifying the Biophysical Properties of Red Blood Cells in Gaucher Disease},
            pdfauthor={Zhaojie Chai; Marine de Person; Pierre A. Buffet; Melanie Franco; George Em Karniadakis}}

\title{Quantifying the Biophysical Properties of Red Blood Cells in Gaucher Disease}

\author[1,*]{Zhaojie Chai}
\author[5]{Marine de Person}
\author[2,4]{Pierre A. Buffet}
\author[3,4,*]{Melanie Franco}
\author[1,*]{George Em Karniadakis}
\runningtitle{Biophysics of RBCs in Gaucher disease} 
\runningauthor{Chai et al.} 

\affil[1]{Division of Applied Mathematics, Brown University, Providence, Rhode Island, United States}
\affil[2]{Université Paris Cité, INSERM, EFS, BIGR U1134, Team BioTiGR, 75015 Paris, France}
\affil[3]{Université Paris Cité, INSERM, EFS, BIGR U1134, Team PAMS, 75015 Paris, France}
\affil[4]{Initiatives IdEx Globule Rouge d'Excellence (InIdex GR-Ex), Université Paris Cité, Paris, France}
\affil[5]{Institut de Chimie Physique, CNRS UMR 8000, Université Paris Saclay, Orsay Cedex 91405, France}

\corrauthor[*]{zhaojie\_chai@brown.edu(ZC); melanie.franco@inserm.fr(MF); george\_karniadakis@brown.edu(GEK)}

\papertype{Article}

\begin{document}

\begin{frontmatter}

\begin{abstract}
Gaucher disease (GD), the most common lysosomal storage disorder, alters red blood cell (RBC) mechanics and circulation, contributing to vascular occlusions, bone infarcts, and splenomegaly. However, the individual roles of GD-RBC biophysical properties in these processes remain unclear. Here, we present a combined computational–experimental investigation to quantitatively characterize GD-RBC biophysical properties and determine how specific mechanical parameters drive abnormal RBC behavior. Informed by experimental data, we independently quantify key RBC properties, including shear modulus ($\mu$), surface-to-volume ratio ($S/V$), and bending modulus ($k_c$). Based on these parameters, we construct three GD-RBC subtypes (GD-RBC1–3) to systematically isolate their individual contributions. At the single-cell level, optical tweezers simulations show up to $\sim$27\% reduction in axial diameter and $\sim$42\% reduction in transverse compression. Tank-treading dynamics exhibit non-monotonic behavior, with rotation frequencies increasing by up to $\sim$70\% or decreasing under elevated bending rigidity. In confined flow, traversal times through microchannel constrictions increase by more than a factor of two, while splenic slit passage times rise from $\sim$250~ms (control) to $>1200$~ms for the severe GD-RBC subtype, approaching a functional no-passage threshold. At the population level, viscosity simulations demonstrate that these alterations collectively elevate blood viscosity, with small fractions ($\sim$4.0\%) of highly rigid cells disproportionately increasing flow resistance. Overall, this study provides a quantitative and mechanistic framework that disentangles the contributions of key RBC parameters to abnormal behavior in GD, linking cellular-scale biophysics to hematologic dysfunction and microvascular occlusion.
\end{abstract}

\begin{sigstatement}
Gaucher disease alters red blood cell mechanics, leading to impaired deformability, abnormal geometry, and enhanced aggregation. Using validated dissipative particle dynamics simulations together with experimental data, we show how these alterations impair RBC behavior at both the single-cell and suspension levels. Our results demonstrate that even a small fraction of highly rigid GD-RBCs disproportionately elevates blood viscosity and obstructs passage through splenic slits, providing a mechanistic explanation for splenomegaly and vascular complications observed in patients. These findings highlight RBC deformability, bending modulus, and geometry as critical biophysical factors in GD, and suggest that targeting these properties could improve microcirculatory flow. This work exemplifies how computational modeling can bridge cellular-scale mechanics with clinical manifestations in hematologic disorders.
\end{sigstatement}

\end{frontmatter}

\section*{Introduction}
Red blood cells (RBCs) are essential for efficient oxygen delivery, relying on remarkable deformability to squeeze through microvessels and splenic slits often narrower than the cell’s resting diameter~\cite{Mebius2005Structure}. This extreme flexibility---combined with the cell’s low internal viscosity and minimal aggregation---ensures uniform tissue perfusion and stable blood rheology~\cite{Chien1970Shear,baskurt2003blood}. In normal blood, high RBC deformability aids flow in both large vessels and capillaries, while controlled, reversible aggregation at low shear helps modulate viscosity~\cite{baskurt2003blood}. By contrast, any impairment in RBC deformability or abnormal increase in aggregation can disrupt microcirculatory flow, elevate blood viscosity, and ultimately compromise oxygen delivery. Such hemorheological disturbances are implicated in the vascular complications of various diseases~\cite{baskurt2003blood}.

Gaucher disease (GD) is the most common lysosomal storage disorder, caused by an inherited deficiency of the enzyme glucocerebrosidase. This enzymatic defect leads to accumulation of sphingolipids in various cells and tissues~\cite{Stirnemann2017Pathophysiology,daho2026effect}. GD is classically characterized by lipid-laden macrophages in the liver, spleen, and bone marrow, resulting in organomegaly and bone marrow dysfunction~\cite{Degnan2019Imaging}. However, there is growing evidence that RBC abnormalities play a significant role in GD pathophysiology alongside macrophage dysfunction~\cite{franco2013abnormal}. RBCs from GD patients are less deformable and exhibit morphological anomalies compared to healthy RBCs~\cite{franco2013abnormal}. GD-RBCs also exhibit a higher aggregation property, an increased disaggregation threshold, and elevated blood viscosity under low-shear conditions~\cite{adar2006aggregation,adar2008increased}. Notably, adhesive interactions are enhanced: under flow, GD-RBCs adhere more strongly to endothelial cells and extracellular matrix components than normal RBCs, partly due to overexpression and hyperactivation of the laminin-binding receptor~\cite{franco2013abnormal}. This combination of stiffer, stickier, and more aggregation-prone RBCs can impair microcirculatory flow and is thought to underlie ischemic complications observed in GD, including the high incidence of bone infarcts and avascular necrosis~\cite{hughes2019gaucher, lafforgue2006pathophysiology, chai2025silico}.

Emerging research suggests that many of these RBC abnormalities in GD arise from altered membrane composition due to sphingolipid overload~\cite{dupuis2020effects}. Even though mature erythrocytes lack lysosomes, studies show that RBCs in GD accumulate excess glucosylceramide and related sphingolipids, both during erythropoiesis and from plasma~\cite{dupuis2020effects}. This abnormal lipid incorporation correlates with impaired cell deformability and hyperaggregation~\cite{dupuis2020effects}. Mechanistically, excess membrane sphingolipids can increase bilayer rigidity and alter protein distribution, potentially disrupting membrane-cytoskeleton coupling~\cite{franco2013abnormal}. Franco et al.~\cite{franco2013abnormal} observed that GD-RBCs form elongated membrane tethers under modest shear stress---suggesting a weakening of the vertical interactions between bilayer and spectrin cytoskeleton---a behavior not seen in healthy cells. Such weakened membrane–skeleton coupling promotes microvesiculation and loss of surface area, driving RBCs toward a more spheroidal, less deformable shape~\cite{dupuis2020effects,deplaine2011sensing,li2015vesiculation,li2016modeling,chai2025silico}.

Despite these insights, the mechanistic link between GD-induced RBC alterations and hemorheological consequences remains incompletely understood. Clinical studies highlight abnormal deformability, altered geometry, and hyperaggregation, but in vitro approaches cannot isolate their individual contributions, especially given the small patient cohorts and significant variability across samples~\cite{franco2013abnormal,dupuis2020effects}. A quantitative framework that integrates these factors in physiologically relevant flow conditions has been lacking.

Here, we develop a dissipative particle dynamics (DPD) framework to systematically quantify how GD-associated changes in RBC biophysics impact microcirculatory transit and vascular occlusions. Informed by experimental data and prior studies, we focus on three key biophysical properties: increased shear modulus, reduced surface-to-volume ratio, and elevated bending modulus. Based on these parameters, we construct three GD-RBC subtypes (GD-RBC1–3) to systematically incorporate and isolate the effects of these biophysical alterations. Using this framework, we investigate both single-cell and suspension-level phenomena. At the single-cell level, we quantify morphological alterations using static DPD simulations and characterize static deformation through optical tweezers simulations. Dynamic membrane behavior is further assessed via tank-treading simulations under shear flow. To evaluate microcirculatory impairment, we quantify RBC passage delays with simulations in confined geometries, including microchannel constrictions and splenic slit traversal. At the suspension level, we compute shear-dependent blood viscosity for mixed RBC populations that reflect experimentally observed subtype distributions, enabling direct comparison with experimental measurements of patient blood. Together, these simulations bridge experimental observations with mechanistic insight. By integrating computational modeling with experimental data, this study provides a comprehensive framework for understanding the hemorheological consequences of RBC alterations in Gaucher disease.

\section*{Materials and Methods}
\subsection*{Experiment setup}
To validate and inform our simulations, we used the elongation-index (EI) measurements of the Dupuis et al.~\cite{dupuis2020effects,dupuis2022phagocytosis} cohort. RBC deformability was assessed on fresh blood samples at 37$^\circ$C across nine shear stresses ranging from 0.30 to 30~Pa by laser diffraction analysis (ektacytometry) on the laser-assisted optical rotational cell analyzer (LORCA, RR Mechatronics). Table~\ref{tab:deformability} summarises the EI values at each shear stress in control subjects (CTR; $n=7$) and untreated Gaucher disease patients (GD; $n=15$), reported as mean $\pm$ standard deviation. In control subjects the EI increases monotonically with shear stress, from $0.08 \pm 0.08$ at 0.30~Pa to $0.58 \pm 0.01$ at 30.00~Pa. GD patients exhibit systematically lower EI values than controls at low shear stress, with statistically significant differences at 0.53~Pa and 0.95~Pa (Welch's $t$-test, $p<0.05$). The differences diminish with increasing stress and, from 1.69~Pa upward, the GD and CTR values are statistically indistinguishable. Consistent with prior reports~\cite{franco2013abnormal,dupuis2020effects}, GD-RBCs are therefore significantly less deformable than normal RBCs specifically under low shear stress.

\begin{table}[htbp]
\centering
\caption{RBC deformability (elongation index, EI) at nine shear stresses (Pa) in control (CTR) subjects and untreated Gaucher disease (GD) patients, from the Dupuis et al.~\cite{dupuis2020effects,dupuis2022phagocytosis} cohort.}
\label{tab:deformability}
\small
\setlength{\tabcolsep}{4pt}
\begin{tabular}{lccccccccc}
\hline
 & \textbf{0.30} & \textbf{0.53} & \textbf{0.95} & \textbf{1.69} & \textbf{3.00} & \textbf{5.33} & \textbf{9.49} & \textbf{16.87} & \textbf{30.00}\\
\hline
CTR (EI) & 0.08 $\pm$ 0.08 & 0.10 $\pm$ 0.02 & 0.17 $\pm$ 0.02 & 0.26 $\pm$ 0.02 & 0.35 $\pm$ 0.02 & 0.44 $\pm$ 0.02 & 0.50 $\pm$ 0.02 & 0.55 $\pm$ 0.01 & 0.58 $\pm$ 0.01 \\
GD (EI)  & 0.06 $\pm$ 0.07 & 0.08 $\pm$ 0.02* & 0.15 $\pm$ 0.02* & 0.24 $\pm$ 0.03 & 0.34 $\pm$ 0.03 & 0.42 $\pm$ 0.04 & 0.49 $\pm$ 0.03 & 0.54 $\pm$ 0.03 & 0.58 $\pm$ 0.03 \\
\hline
\end{tabular}

\begin{minipage}{14cm}
\vspace{0.1cm}
\small Notes: Values are mean $\pm$ standard deviation. Group comparison by Welch's $t$-test (*$p<0.05$). CTR: $n=7$; GD: $n=15$. EI: elongation index.
\end{minipage}
\end{table}

The shear stress level of $\sim$3~Pa is conventionally regarded as the threshold distinguishing low-to-moderate from high shear stress conditions. Below 3~Pa, RBC deformability primarily depends on the ability of the RBC membrane to deform under shear. In contrast, beyond 3~Pa, deformability predominantly reflects the internal viscosity of the cells. Upon examination of the referenced guidelines~\cite{baskurt2009guidelines}, it is apparent that deformability is intricately connected to the viscoelastic properties of the cell membrane. Furthermore, the elongation-deformation research by Tsubota included additional parameters such as elongation and inclination angle, both of which are modulated by membrane viscoelasticity and inner viscosity~\cite{TSUBOTA2021Elongation}. Analysing the data in Table~\ref{tab:deformability}, we note that the EI values for GD and CTR become statistically indistinguishable at shear stresses $\gtrsim$1.69~Pa. This observation, combined with the fact that RBC deformability at elevated stress primarily reflects membrane viscoelasticity and internal viscosity, suggests that the membrane viscoelasticity of GD-RBCs and CTR-RBCs is comparable at high shear, with the principal deformability deficit of GD-RBCs appearing at low shear stress.

To characterise alterations in RBC sphingolipid metabolism, we use the lipidomic dataset of Dupuis et al.~\cite{dupuis2020effects,dupuis2022phagocytosis}, obtained with the UPLC--MS/MS protocol of Chipeaux et al.~\cite{chipeaux2017optimization}. This protocol simultaneously quantifies glucosylceramide (GL-1), glucosylsphingosine (Lyso-GL1), sphingosine (Sph), and sphingosine-1-phosphate (S1P) in plasma and RBCs, with concentrations expressed as nmol/L of whole blood by accounting for each subject's hematocrit. The corresponding per-subject sphingolipid levels for controls (\(n=11\)) and untreated GD patients (\(n=16\)) used in this work are summarised in Table~\ref{tab:sphingolipid_levels}.

\subsection*{Parameter setup for Gaucher Disease (GD) RBC models}

Both the patient-level lipidomic measurements (Table~\ref{tab:sphingolipid_levels})~\cite{dupuis2020effects,dupuis2022phagocytosis} and our computational results in the Supporting Material (Fig.~S2) consistently demonstrate pronounced enrichment of sphingolipids in GD-RBC membranes, including elevated levels of glucosylceramide, lyso-GL1, sphingosine, and S1P. Model membrane studies further support that changes in lipid composition, including glycosphingolipid or cholesterol enrichment, substantially modulate bending rigidity and can increase the effective bending modulus two- to three-fold~\cite{bochicchio2016membrane, Khmelinskii2020Swelling,li2014erythrocyte}. Taken together, these results provide the biophysical basis for assigning an elevated bending modulus ($k_c=4.8 \times 10^{-19}$~J) in the GD-RBC3 model (Table~\ref{tab:mechanical_properties}), linking sphingolipid enrichment to altered membrane curvature and reduced deformability in Gaucher RBCs.


\begin{table}[ht]
\centering
\caption{Sphingolipid levels in RBCs of control ($n=11$) and untreated GD ($n=16$) subjects, from the Dupuis et al.~\cite{dupuis2020effects,dupuis2022phagocytosis} cohort (quantification method: Chipeaux et al.~\cite{chipeaux2017optimization}). Values are in nmol/L of whole blood.}
\label{tab:sphingolipid_levels}
\small
\begin{tabular}{l c c c c}
\hline
Group & GL-1 & Lyso-GL1 & Sph & S1P \\
\hline
CTR & 120.25 &  5.44 & 1.63 & 1{,}053.94 \\
CTR &  57.67 &  5.20 & 0.95 & 1{,}164.57 \\
CTR &  54.93 &  5.98 & 4.00 & 1{,}330.29 \\
CTR &  69.00 &  4.84 & 2.21 &    935.63 \\
CTR &  68.02 &  5.82 & 4.57 & 1{,}473.63 \\
CTR &  82.36 &  4.92 & 4.09 & 1{,}337.20 \\
CTR &  82.94 &  5.80 & 4.00 & 1{,}629.01 \\
CTR & 131.19 &  5.21 & 1.17 & 1{,}563.42 \\
CTR &  93.34 &  5.05 & 5.01 & 1{,}098.66 \\
CTR &  83.55 &  5.15 & 4.70 & 1{,}149.52 \\
CTR &  90.56 &  5.46 & 4.72 & 1{,}274.03 \\
GD  & 146.10 &  64.95 &  8.92 & 2{,}583.61 \\
GD  &  97.17 &  54.93 & 11.02 & 2{,}817.87 \\
GD  & 108.42 &  82.60 &  8.31 & 2{,}639.82 \\
GD  & 171.37 &  39.77 & 10.64 & 2{,}165.20 \\
GD  & 147.14 &  86.98 & 10.21 & 2{,}640.05 \\
GD  & 116.03 &  70.86 & 10.21 & 3{,}032.91 \\
GD  & 204.45 &  78.52 & 10.15 & 3{,}016.05 \\
GD  &  85.88 &  51.49 &  6.66 & 2{,}214.62 \\
GD  &  63.24 & 184.45 & 12.53 & 3{,}041.77 \\
GD  & 184.08 &  59.42 & 10.27 & 2{,}984.12 \\
GD  &  64.56 &  90.35 & 13.28 & 4{,}071.77 \\
GD  & 108.19 & 164.93 & 16.84 & 2{,}885.60 \\
GD  & 153.98 &  87.84 & 21.15 & 2{,}942.90 \\
GD  & 121.87 &  64.69 &  8.44 & 2{,}503.87 \\
GD  & 113.70 &  65.42 &  8.50 & 2{,}443.43 \\
GD  & 163.92 &  38.85 & 16.31 & 1{,}861.51 \\
\hline
\textbf{Controls (mean $\pm$ SD)} & \textbf{84.89 $\pm$ 23.84} & \textbf{5.35 $\pm$ 0.38} & \textbf{3.37 $\pm$ 1.55} & \textbf{1{,}273.63 $\pm$ 218.62} \\
\textbf{GD patients (mean $\pm$ SD)} & \textbf{128.13 $\pm$ 41.34} & \textbf{80.38 $\pm$ 40.23} & \textbf{11.46 $\pm$ 3.79} & \textbf{2{,}740.32 $\pm$ 498.07} \\
\hline
\end{tabular}
\end{table}

The decrease in RBC deformability below 3 Pa appears to primarily reflect a loss of membrane flexibility \cite{Renoux2019Impact}. This loss of membrane flexibility is associated with alterations in the bending modulus of the membrane \cite{bochicchio2016membrane, Khmelinskii2020Swelling, li2014erythrocyte, chai2022periodic, chai2023dynamics}. In addition to reduced flexibility, the GD-RBC3 model further assumes an increase in the membrane bending modulus by a factor of two compared to normal RBCs. Table~\ref{tab:mechanical_properties} summarizes the model parameters for RBCs under normal and Gaucher disease (GD) conditions, including shear modulus ($\mu$), surface-to-volume ratio ($S/V$), and bending modulus ($k_c$), which together govern RBC mechanics and deformability. The shear modulus $\mu$ characterizes in-plane membrane elasticity and resistance to stretching, and was calibrated by matching elongation under shear stress with experimental data~\cite{chai2025silico}. The $S/V$ ratio determines geometric deformability and the ability of RBCs to traverse narrow constrictions, and was obtained from projected area measurements consistent with imaging data. The bending modulus $k_c$ governs out-of-plane rigidity, particularly relevant for high-curvature deformations such as slit traversal, and was guided by lipid enrichment experiments~\cite{franco2013abnormal,dupuis2020effects}. Together, these parameters enable systematic dissection of their individual and combined contributions to RBC biomechanics.

Before introducing the three GD-RBC variants, we emphasise that the terms ``mild'' and ``severe'' used throughout refer to the level of biophysical and morphological alteration captured by our RBC subtype models (GD-RBC1, GD-RBC2, GD-RBC3); they do not correspond to clinical disease-severity scoring, which is established by physicians from distinct hematological and biological markers. Biologically, a GD blood sample contains a heterogeneous mixture of young and old RBCs, with normal and abnormal morphologies coexisting in the same circulation. Because RBCs accumulate sphingolipids progressively during their lifetime in plasma, older cells tend to carry a higher lipid burden and are more likely to display abnormal morphology and reduced deformability~\cite{dupuis2020effects,dupuis2022phagocytosis}. Our GD-RBC1--3 variants are designed to capture, in a categorical fashion, distinct points along this continuum of cellular ageing and lipid-driven membrane alteration.

Notably, under GD conditions the shear modulus of RBCs (GD-RBC1--3) is set to nine times the control value, $\mu = 9\,E_{s0} = 42.57~\mu\mathrm{N/m}$, where $E_{s0} = 4.73~\mu\mathrm{N/m}$ is the CTR-RBC shear modulus---a nine-fold elevation that indicates pronounced membrane stiffening, consistent with the calibration in our previous work~\cite{chai2025silico}. In GD-RBC1, only the shear modulus is elevated, while the surface-to-volume ratio ($S/V=1.44$) and bending modulus ($k_c=2.4\times10^{-19}$~J) remain unchanged, isolating the effect of membrane stiffening. In GD-RBC2, the $S/V$ ratio is reduced to 1.22 with the same bending modulus, introducing geometric alterations that further impair deformability. In GD-RBC3, both reduced $S/V$ (1.22) and increased bending modulus ($k_c=4.8\times10^{-19}$~J) are incorporated, representing additional resistance to curvature deformation. We elevate $k_c$ only in GD-RBC3 because this subtype is intended to represent the most severely lipid-enriched subpopulation: the per-subject lipidomic data (Table~\ref{tab:sphingolipid_levels}) and our coarse-grained molecular dynamics (CGMD) simulations (Fig.~S2) indicate that membrane curvature elasticity is appreciably altered only above a critical sphingolipid enrichment level (here $R$ denotes the mole fraction of sphingolipid-like coarse-grained particles in the model bilayer; details in SI), with a critical value $R\approx 0.35$ that is reached in the most-affected fraction of GD-RBCs but not necessarily in milder cases captured by GD-RBC1 and GD-RBC2. For controlled comparison, $S/V$ is fixed at 1.22 in both GD-RBC2 and GD-RBC3, such that the only difference between these two models is the bending modulus. Taken together, these parameter sets highlight that GD-RBCs exhibit heterogeneity in their mechanical characteristics: GD-RBC1 represents primarily stiffened membranes, GD-RBC2 incorporates altered cell geometry, and GD-RBC3 manifests both geometric and bending rigidity changes. Such distinctions are crucial for understanding the diverse biomechanical manifestations of GD pathology.

\begin{table}[htbp]
\centering
\caption{Mechanical properties of normal and GD-RBC models.}
\label{tab:mechanical_properties}
\begin{tabular}{lcccc}
\hline
 & \textbf{CTR-RBC} & \textbf{GD-RBC1} & \textbf{GD-RBC2} & \textbf{GD-RBC3} \\
\hline
Shear modulus ~\cite{chai2025silico}. $\mu$ (\(\mu\mathrm{N/m}\)) & 4.73 & 42.57 & 42.57 & 42.57 \\
Surface-to-volume ratio $S/V$ ($\mu$m$^{-1}$) & 1.44 & 1.44 & 1.22 & 1.22 \\
Bending modulus $k_{c}$ ($10^{-19}$~J) & 2.4 & 2.4 & 2.4 & 4.8 \\
\hline
\end{tabular}
\end{table}

\subsection*{Dissipative particle dynamics (DPD) model}

We employed the DPD method to simulate the blood plasma as a mesoscopic fluid, following the framework established by Groot and Warren~\cite{groot1997dissipative} and Hoogerbrugge and Koelman~\cite{hoogerbrugge1992simulating}. In this method, the plasma is represented by DPD particles that interact through soft conservative, dissipative, and random forces, which collectively enforce hydrodynamic behavior \cite{Fedosov2011Multiscale}. The dissipative and random forces satisfy the fluctuation--dissipation theorem, ensuring proper thermal equilibration of the system \cite{groot1997dissipative}. The fluid viscosity was calibrated by measuring the shear stress response under steady shear flow, and parameters were chosen to reproduce the physiological viscosity of blood plasma while maintaining numerical stability~\cite{Fedosov2011Quantifying}. In our simulations, the viscosity of the cytoplasm was assumed equal to that of the plasma, which has been shown to adequately capture suspension rheology. Similar DPD parameter choices have been shown in prior work to accurately predict blood viscosity over a range of shear rates~\cite{Fedosov2011Predicting}. The specific DPD parameters, including number density, conservative force coefficient, dissipative force constant, and time step, are listed in Table~S1 of the Supporting Material. The computational domain was periodic in the flow and vorticity directions, with no-slip walls moving at constant velocities to impose shear flow in the viscosity simulations. For confined flow simulations, a constant pressure gradient was applied to drive RBCs through capillary-like channels or splenic slit geometries. All simulations were run for sufficient time to reach steady-state after initial transients.

\subsection*{RBC model}

To investigate the role of each biophysical factor on the blood viscosity and, ultimately, GD-RBC computationally, we need to model the individual RBCs accurately, without compromising the essential hydrodynamics of the system and the convective transport processes that govern the blood flow. A normal RBC is a highly deformable, nucleus-free biconcave cell with a resting diameter of $\sim$8~$\mu$m, and is well described as a viscoelastic object. This deformability allows RBCs to squeeze through capillaries as small as 3~$\mu$m, preserve their shape at small deformation rates, and orient along the flow direction at larger deformation rates in wider arteries. The RBC membrane is modeled as a set of $N_{\nu}$ DPD particles whose three-dimensional coordinates $\mathbf{X}_i$ ($i = 1,\ldots, N_{\nu}$) form a triangulated spring network, with each edge carrying a dashpot to model membrane viscosity. To model the incompressibility of the RBC membrane, area and volume constraints are applied. In addition, considering the bending resistance between all neighboring triangles, the bending rigidity of the membrane can be mimicked. The shear modulus and bending modulus of normal RBCs are taken as $E_{s0} = 4.73~\mu\text{N}/\text{m}$ and $E_{b0} = 2.4 \times 10^{-19}~\text{J}$, respectively, while the membrane viscosity is $\eta_{m} = 0.128~\text{Pa}\cdot\text{s}$ \cite{fedosov2010multiscale}.

\subsection*{Cell-cell interaction models}

In our simulations, we model cell–cell adhesion by introducing an attractive potential between RBC membranes. Specifically, an attractive Morse potential is applied between certain membrane particles on different RBCs, following a similar approach used in prior studies of RBC aggregation \cite{franco2013abnormal,chai2025silico}. These interactions are approximated with the Morse potential, defined as
\begin{equation}
V(r) = D_e \left( e^{-2\beta(r - r_0)} - 2 e^{-\beta(r - r_0)} \right),
\end{equation}
where $r$ denotes the distance between two particles, $D_e$ is the depth of the potential well, $\beta$ denotes the interaction range, and $r_0$ represents the zero-force distance. The Morse potential is applied to a specific type of vertices of each RBC, which are called “interactive vertices”, characterizing two severity levels, severe and mild, as studied in \cite{deng2020quantifying}. In addition, to prevent RBC membranes from overlapping, we applied a repulsive term of Lennard-Jones potential to all membrane vertices \cite{fedosov2010multiscale}; this potential is given by

\begin{equation}
U(r) =
\begin{cases}
4\epsilon \left[ \left( \dfrac{\sigma}{r} \right)^{12} - \left( \dfrac{\sigma}{r} \right)^6 \right] , & r \leq 2^{1/6}\sigma, \\
0, & r > 2^{1/6}\sigma,
\end{cases}
\end{equation}

where $\epsilon$ and $\sigma$ are scaling constants for energy and distance, respectively, and these interactions vanish for $r > 2^{1/6}\sigma$.

\subsection*{Computational framework}

In this study, we employ a multiscale simulation approach to capture both single-cell biomechanics of RBCs and their suspension rheology under GD conditions. At the mesoscopic scale, we use the DPD method, which is widely applied to blood cell and soft matter systems~\cite{fedosov2010multiscale,li2014erythrocyte}. In the DPD framework, particles interact through conservative, dissipative, and random forces, the latter two forming a thermostat consistent with the fluctuation–dissipation theorem~\cite{groot1997dissipative}. This approach enables computationally efficient simulations of RBC mechanics and blood flow while preserving essential hydrodynamic interactions. Alongside such physics-based formulations, data-driven, deep-reinforcement-learning, and agentic self-improving strategies have been developed for particle-based systems and for the autonomous discovery of numerical algorithms~\cite{sanchez2020learning,zhang2020deep,zhang2021deep,toscano2026graft}, highlighting the breadth of computational tools now available for mesoscopic simulations. Details of the DPD implementation and parameter choices are provided in the Supporting Material.  

Additional molecular-scale CGMD simulations of the RBC membrane are presented in the Supporting Material to complement the DPD framework \cite{chai2023dynamics}; the CGMD parameters are listed in Table~S2. These simulations resolve membrane-level changes associated with sphingolipid enrichment in Gaucher RBCs, thereby providing mechanistic insights beyond the mesoscopic description; confocal images of vesiculating GD-RBCs from a patient sample are shown in Fig.~S1.

\subsection*{Simulation setup}

Consistent with the design of the microfluidic experiments, each simulation was specified by defining the computational domain, boundary conditions, particle resolution, and flow-driving mechanism, as described below.

\textit{Optical tweezers test.} 
Optical tweezers simulations are used to quantify single-cell deformability and extract membrane mechanical properties, particularly the effective shear modulus and elastic response under controlled tensile loading. These tests provide a direct measure of RBC stiffness by relating applied force to axial and transverse deformation \cite{dao2003mechanics}. Static deformation tests were performed on a single RBC model ($N_\nu=500$ membrane vertices) suspended in a cubic box of $15 \times 15 \times 15~\mu$m$^3$ with periodic boundaries. Stretching was imposed by tethering two diametrically opposite vertices to harmonic springs and applying constant tensile forces over the range 0--200~pN; the snapshots in Fig.~\ref{fig:stretching}A correspond to 100~pN.

\textit{Tank-treading dynamics.} 
Tank-treading simulations characterize dynamic membrane response under shear flow, providing information on membrane viscosity, shear elasticity, and the transition between dynamical regimes. The tank-treading frequency serves as a sensitive indicator of RBC mechanical integrity and flow adaptability \cite{fedosov2010multiscale}. Tank-treading motion was simulated by placing a single RBC in a shear flow domain of $20 \times 20 \times 20~\mu$m$^3$, bounded by walls of thickness $5~\mu$m. Shear flow was generated by moving the bounding walls in opposite directions at constant velocity. Periodic boundary conditions were imposed along the flow ($x$) and vorticity ($y$) directions, while no-slip conditions were applied along the gradient ($z$) direction. Shear rates ranged from $50$ to $250$~s$^{-1}$, and the angular displacement of a tagged membrane vertex was tracked to determine the tank-treading period.

\textit{Microchannel constriction.} 
Microchannel constriction simulations probe the ability of RBCs to deform and traverse narrow capillary-like geometries, providing a direct measure of microcirculatory transit efficiency and flow resistance. Metrics such as traversal time and velocity reflect the combined effects of membrane stiffness and cell geometry \cite{liu2021computational}. The constriction was modeled using a rectangular channel of $60 \times 10 \times 10~\mu$m$^3$ with a narrow section ($30~\mu$m in length, $5~\mu$m in width, $2.7~\mu$m in depth). A pressure gradient of $\Delta P=6$--12~Pa$\cdot\mu$m$^{-1}$ drove the flow. Each simulation initialized a single RBC 5~$\mu$m upstream of the constriction entrance, with periodic boundaries in the flow direction and no-slip wall conditions.

\textit{Splenic slit traversal.} 
Splenic slit simulations mimic the biomechanical filtration function of the spleen, where RBCs must undergo extreme deformation to pass through interendothelial slits. Passage time and failure to traverse serve as indicators of splenic retention and clearance susceptibility \cite{du2026high}. The slit geometry consisted of a height of $1.2~\mu$m, width of $5.0~\mu$m, and depth of $2.5~\mu$m within a channel of $30 \times 10 \times 10~\mu$m$^3$. A single RBC was positioned 9~$\mu$m upstream, and a constant pressure gradient of $6$~Pa$\cdot\mu$m$^{-1}$ was applied. Passage time and deformation were recorded.

\textit{Viscosity simulations.} 
Viscosity simulations quantify bulk hemorheological properties, linking single-cell mechanics to macroscopic flow behavior. Shear-dependent viscosity reflects collective effects of deformability, cell-cell interactions, and population heterogeneity \cite{Fedosov2011Predicting}. Viscosity was examined in a domain of $60 \times 60 \times 50~\mu$m$^3$. Shear flow was generated by translating bounding walls, with periodic boundaries in the flow ($x$) and vorticity ($z$) directions. A total of 558 RBCs ($N_\nu=500$ each) were simulated to achieve a hematocrit of $H_t=36\%$, with approximately 600,000 plasma particles, yielding $\sim$879,000 particles in total.

Across all simulations, system sizes ranged from $1.2\times10^4$ particles (single-cell tests) to nearly $9\times10^5$ particles (viscosity simulations).  All simulations were performed using an extended in-house version of the LAMMPS (Large-scale Atomic/Molecular Massively Parallel Simulator) code. Each simulation took approximately $1\times10^6$ to $2\times10^6$ time steps. A typical simulation requires 1200 CPU core hours to 2400 CPU core hours by using the computational resources (Intel Xeon E5-2670 2.6 GHz 24-core processors) at the Center for Computation and Visualization at Brown University.

 \begin{figure}[!tbp]
\begin{center}
\includegraphics[width=1.000\textwidth]{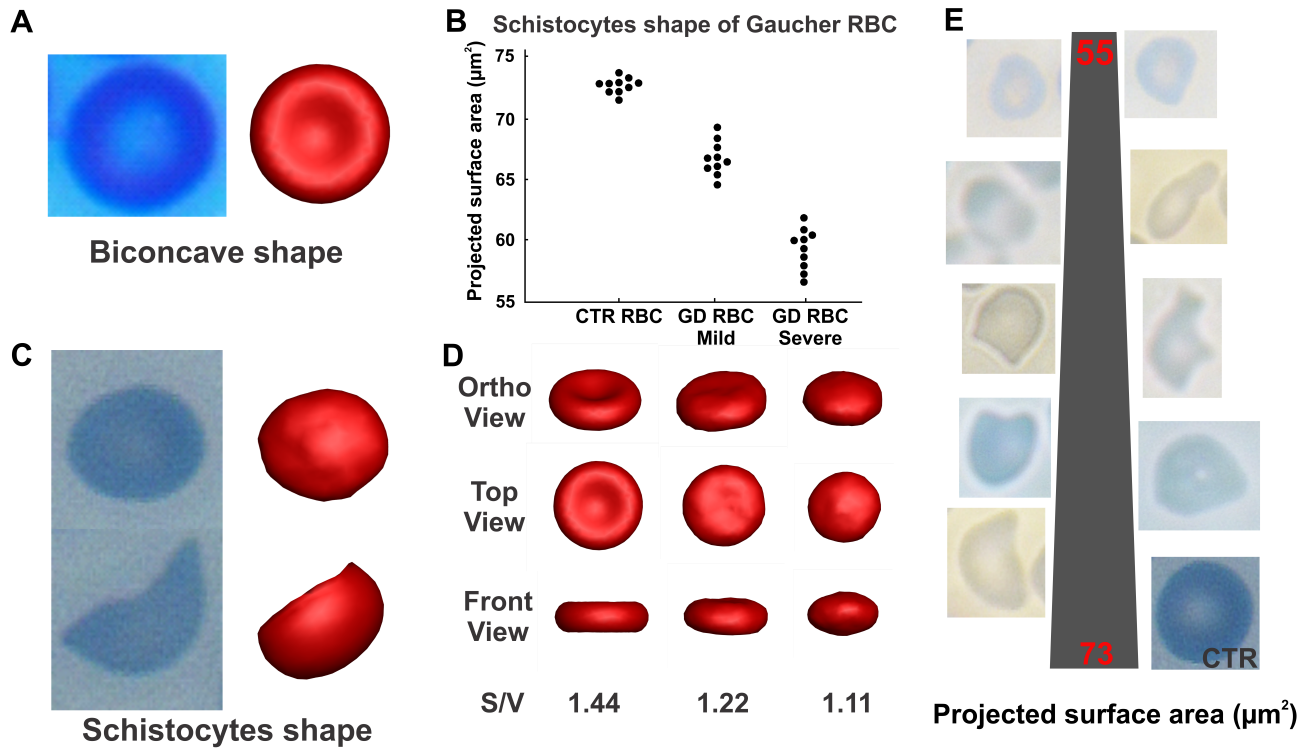}
\end{center}
\vspace{-0.15in}
\caption{\small{\bf Morphological characteristics and surface area-to-volume (S/V) ratios of GD-RBCs and controls.} (A) control RBC (CTR-RBC) displaying a normal biconcave shape.
(B) Projected surface area measurements for CTR-RBC and RBCs from GD patients, categorized by mild and severe RBC-morphology classes (\textit{not} clinical disease severity). The data indicate reduced surface area in GD-RBCs, particularly in the morphologically most affected cases, reflecting RBC-level alterations associated with the disease.
(C) Schistocyte shapes observed in GD, highlighting the irregular, fragmented morphology in contrast to the typical biconcave shape of healthy RBCs.
(D) Orthogonal, top, and front views of simulated RBC models with different S/V ratios: 1.44 for CTR-RBC, 1.22 for GD-RBC (mild), and 1.11 for GD-RBC (severe). These illustrate progressive shape changes and reduced S/V ratios in GD-RBCs. 
(E) Representative images of RBCs sorted by decreasing projected surface area, from 73–55~$\mu$m$^2$, including typical schistocyte morphologies observed in GD. Reduced surface area and irregular shapes further support the presence of altered membrane geometry in GD-RBCs. CTR-RBCs (n = 10) are from 7 healthy volunteers; GD-RBCs (n = 20) are from 9 GD patients. \textit{Note:} The model with $S/V=1.11$ in (D) illustrates morphology only; subsequent controlled simulations use $S/V=1.22$ for both GD-RBC2 and GD-RBC3 to isolate bending effects.} 
\label{fig:morphology}
\end{figure}

\section*{Results}

\subsection*{Alterations in Morphology and Deformability of GD-RBCs}

GD-RBCs exhibited pronounced morphological alterations, as demonstrated by experimental observations (Fig.~\ref{fig:morphology}). In control samples, RBCs retained the characteristic biconcave shape (Fig.~\ref{fig:morphology}A), consistent with the baseline model geometry. In contrast, RBCs from GD patients frequently displayed schistocyte-like morphologies—irregular, fragmented, or misshapen cells—indicative of severe membrane remodeling and loss of structural symmetry (Fig.~\ref{fig:morphology}C,E). To quantify these morphological changes, we measured the projected surface area from confocal images (Fig.~\ref{fig:morphology}B,E), revealing a progressive reduction from controls to the mild- and severe-morphology GD-RBC classes (defined here at the RBC level, not by clinical disease severity). CTR-RBCs (n = 10 from 7 healthy volunteers) clustered tightly around 73 $\mu$m$^{2}$ with minimal variability, whereas GD-RBCs (n = 20 from 9 patients) exhibited smaller and more broadly distributed areas, with mild-morphology GD-RBCs averaging $\sim$66 $\mu$m$^{2}$ and severe-morphology GD-RBCs reaching values as low as 55 $\mu$m$^{2}$. The increased scatter in GD groups reflects substantial heterogeneity in cell morphology. Notably, the distributions overlap, indicating the absence of a sharp boundary between CTR, mild-morphology, and severe-morphology GD-RBC populations. For modeling purposes, we therefore introduced categorical distinctions between mild-morphology and severe-morphology GD-RBCs as a practical approximation rather than strict biological thresholds.

We next performed morphology-resolved simulations consistent with these experimental observations. The results show a progressive decrease in the surface-to-volume (S/V) ratio with the morphology class, from 1.44 (CTR) to 1.22 (mild-morphology GD-RBC) and 1.11 (severe-morphology GD-RBC) (Fig.~\ref{fig:morphology}D). As S/V decreases, the canonical biconcave shape becomes progressively less pronounced: top views exhibit reduced diameter, while side views reveal increased thickness. At S/V = 1.11, cells approach a more spheroidal geometry, consistent with impaired deformability. Experimental images sorted by decreasing projected area (73–55 $\mu$m$^{2}$) further illustrate this transition, with CTR-RBCs maintaining uniform biconcavity and GD-RBCs exhibiting marked heterogeneity, including fragmented and schistocyte-like forms (Fig.~\ref{fig:morphology}E). Consistent with these structural changes, elongation index (EI) measurements from ektacytometry confirm that GD-RBCs are significantly less deformable than healthy controls, particularly under low shear stress conditions (Table~\ref{tab:deformability}). Detailed formulations and parameter settings governing membrane mechanics and shape regulation are provided in the Supporting Material.

\subsection*{Optical tweezers test}

\begin{figure}[!tbp]
\begin{center}
\includegraphics[width=1.000\textwidth]{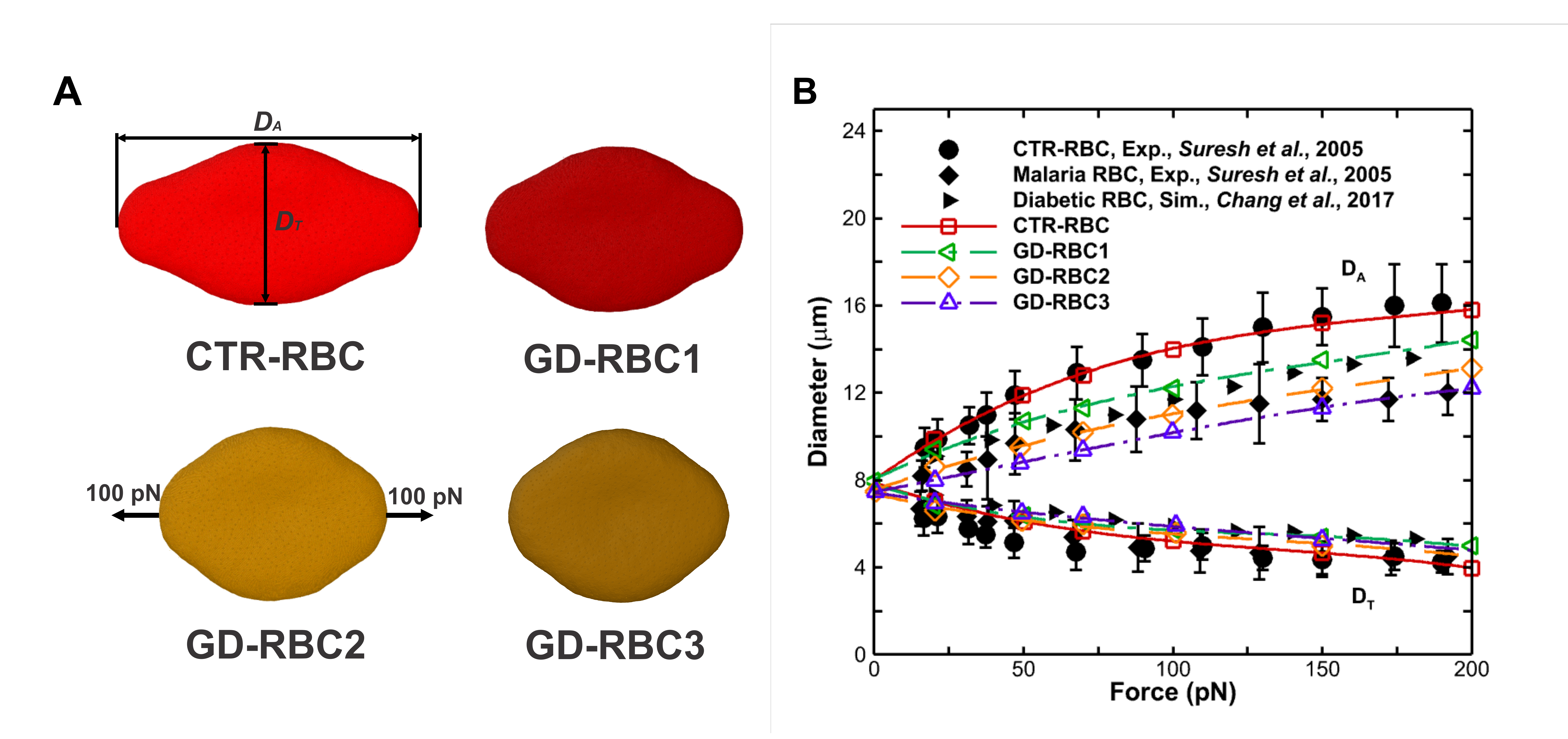}
\end{center}
\vspace{-0.15in}
\caption{\small{\bf Static deformation responses of RBCs tested with in-silico optical tweezers.} (A) Simulation images of RBCs under 100 pN tensile force. CTR-RBCs maintain a larger aspect ratio, with axial diameter (\(D_A\)) and transverse diameter (\(D_T\)), compared to the more compact shapes seen in GD-RBCs (GD-RBC1, GD-RBC2, GD-RBC3), illustrating increased rigidity in GD models. 
(B) Force-diameter relationship for CTR-RBC and GD-RBCs (GD-RBC1, GD-RBC2, GD-RBC3) compared with experimental data for CTR-RBCs and malaria-infected RBCs (Malaria RBC) from Suresh et al.~\cite{suresh2005connections} and diabetic RBC simulations from Chang et al.~\cite{CHANG2017Diabete}. Axial (\(D_A\)) and transverse (\(D_T\)) diameters increase with applied force, with GD-RBCs exhibiting less deformation compared to CTR-RBC, indicating reduced elasticity. The more severe GD-RBC models (GD-RBC2 and GD-RBC3) show significantly lower deformability, consistent with observed rigidity in GD-RBCs.} 
\label{fig:stretching}
\end{figure}

To further evaluate the mechanical response of GD-RBCs, we performed complementary simulations of static stretching under optical tweezers-like tensile loading (Fig.~\ref{fig:stretching}). In this setup, opposing forces were applied along the axial direction to mimic the classical optical trap experiment. At a constant force of $100$~pN, CTR-RBCs displayed pronounced elongation, with the axial diameter ($D_A$) expanding to $\sim$14~$\mu$m and the transverse diameter ($D_T$) contracting to $\sim$5.2~$\mu$m, producing a distinctly elongated morphology (Fig.~\ref{fig:stretching}A). In contrast, GD-RBCs showed progressively reduced deformation: GD-RBC1 reached $D_A \approx 12.2~\mu$m, GD-RBC2 elongated to $\sim$11~$\mu$m, and GD-RBC3 exhibited the stiffest response with $D_A$ barely exceeding 10~$\mu$m while maintaining a comparatively larger $D_T$ ($\sim$5.9~$\mu$m). These values correspond to a $\sim$27\% smaller axial diameter and $\sim$42\% less transverse compression (measured as the change in $D_T$ from the unstressed cell) in GD-RBC3 relative to CTR-RBCs at the same applied force. 

Quantitative analysis of the full force–deformation curves (Fig.~\ref{fig:stretching}B) revealed that $D_A$ increased monotonically with force across all RBC types, but the magnitude of elongation was substantially reduced in GD cells. At 200~pN, CTR-RBCs extended to nearly 16~$\mu$m in $D_A$, whereas GD-RBC2 and GD-RBC3 plateaued at only 12–13~$\mu$m—representing a 17–23\% reduction compared to controls. Similarly, transverse compression ($D_T$ reduction) was attenuated in GD models. CTR-RBCs decreased to $\sim$4.0~$\mu$m, whereas GD-RBC2 and GD-RBC3 remained above 4.5~$\mu$m, despite their smaller initial diameters resulting from membrane loss and reduced $S/V$ ratio. Even after accounting for differences in the original cell shape, GD-RBC2 and GD-RBC3 still exhibited $\sim$10–20\% weaker compressive deformation. Notably, GD-RBC1 showed an intermediate phenotype, falling between the CTR and the more abnormal GD-RBC2/3. When compared against prior experimental and simulation datasets, our findings place GD-RBCs within the spectrum of pathological stiffening. Normal RBCs from Suresh et al.~\cite{suresh2005connections} matched well with our CTR-RBC simulations, while malaria-infected RBCs reported in the same study showed even more severe restriction of elongation than GD-RBCs. Likewise, diabetic RBC simulations from Chang et al.~\cite{CHANG2017Diabete} fell between our CTR and GD cases, further supporting the validity of our parameterization. The degree of stiffening observed in GD-RBC2 and GD-RBC3 approaches the levels reported for malaria-infected RBCs, which can exhibit up to an order-of-magnitude increase in shear modulus. Taken together, these results confirm that GD-RBCs are mechanically compromised, with diminished elasticity and reduced capacity for deformation under tensile stress.

\subsection*{Tank-treading dynamics of RBCs}

\begin{figure}[!tbp]
\begin{center}
\includegraphics[width=1.000\textwidth]{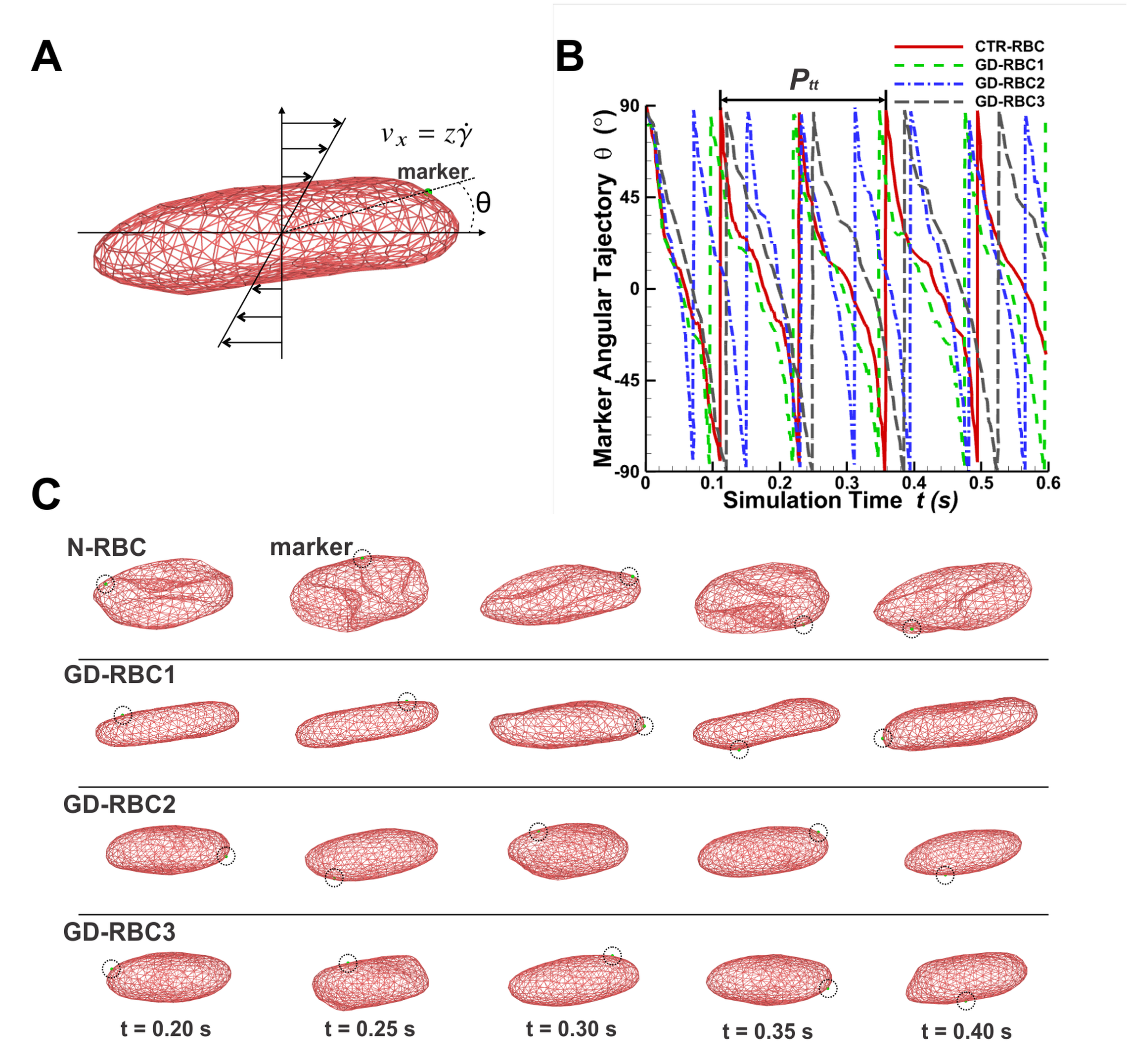}
\end{center}
\vspace{-0.15in}
\caption{\small{\bf Tank-treading dynamics of CTR-RBC and GD-RBCs under shear flow.}  (A) Schematic of tank-treading motion showing the angular trajectory (\(\theta\)) of a marker on the RBC membrane as it moves under shear (\(v_x = z\dot{\gamma}\)). (B) Marker angular trajectories over time for CTR-RBC and GD-RBC models (GD-RBC1, GD-RBC2, GD-RBC3) at \(\dot{\gamma} = 100\)~s\(^{-1}\). Each sawtooth spans a half-rotation of the marker, so the tank-treading period (P\(_{\text{tt}}\)) indicated by the arrow covers two successive sweeps. P\(_{\text{tt}}\) differs by subtype: it is shorter in GD-RBC1/2 (faster rotation) but longer and irregular in GD-RBC3 due to elevated bending rigidity. (C) Snapshots of RBCs showing the tank-treading motion at different time points (t = 0.20 s to 0.40 s) for CTR-RBC (labelled N-RBC in the panel) and GD-RBCs. The position of the marker (circled) highlights the membrane rotation over time. GD-RBC1 and GD-RBC2 exhibit faster tank-treading motion than CTR-RBCs due to their higher shear modulus and lower S/V, while GD-RBC3 shows markedly slower rotation due to increased bending modulus.
} 
\label{fig:tanktreading}
\end{figure}

We next investigated the tank-treading behavior of RBC membranes under shear flow to assess how Gaucher disease alters their dynamic properties. Tank-treading refers to the steady rotational movement of the cell membrane around its internal contents, with the orientation of cells fixed relative to the flow direction. This motion reflects the membrane's ability to continuously deform and adapt to shear, making it a sensitive indicator of mechanical integrity. To quantify these dynamics, we used simulations to track the angular position $\theta(t)$ of a membrane marker over time for control and GD-RBCs subjected to identical shear rates (Fig.~\ref{fig:tanktreading}A). The angular trajectory $\theta(t)$ at $\dot{\gamma} = 100$~s$^{-1}$ is shown in Fig.~\ref{fig:tanktreading}B for the different RBC models, and the TT period $P_{\text{tt}}$, defined as the time required for one full membrane rotation, revealed systematic alterations with increasing GD severity. CTR-RBCs ($P_{\text{tt}} \approx 0.26~s$) exhibited smooth and periodic angular trajectories, reflecting stable TT dynamics and preserved membrane responsiveness. In contrast, GD-RBCs displayed progressively abnormal dynamics. GD-RBC1 ($P_{\text{tt}} \approx 0.23~s$) showed only a mild shortening of the period, largely attributable to its increased shear modulus from $E_{s0}$ to $9E_{s0}$ relative to controls. GD-RBC2 ($P_{\text{tt}} \approx 0.18~s$) rotated at an even faster rate but with irregular periodicity, indicating that altered cell thickness and reduced surface-to-volume ratio can accelerate TT frequency while simultaneously destabilizing the rotation. Strikingly, GD-RBC3 ($P_{\text{tt}} \approx 0.28~s$) exhibited markedly delayed and intermittently interrupted TT motion, suggesting that increased bending rigidity can substantially slow and destabilize the dynamics. These results demonstrate that GD-RBCs do not simply follow a monotonic trend of increasing or decreasing TT frequency; rather, their dynamics reflect the combined and sometimes competing influences of shear modulus, bending rigidity, and $S/V$. While GD-RBC2 transiently accelerates TT as a result of increased shear modulus and reduced $S/V$, GD-RBC3 ultimately suppresses and destabilizes TT motion owing to elevated bending rigidity. Thus, the simulations reveal a spectrum of TT abnormalities in GD-RBCs, highlighting how structural changes at both the membrane and whole-cell levels synergistically impair dynamic deformability under shear~\cite{tran1984determination,fischer2007tank,peng2015stability}.  

Representative snapshots further illustrate these dynamics (Fig.~\ref{fig:tanktreading}C). In CTR-RBCs, the circled marker progresses smoothly along the cell perimeter between $t = 0.20$–$0.40$~s, consistent with stable and periodic TT motion. GD-RBC1 shows only mild deviations, maintaining near-regular trajectories with slightly faster progression. In GD-RBC2, the marker rotates more rapidly but with irregular step-like advances, reflecting destabilized TT dynamics associated with altered cell geometry. By contrast, GD-RBC3 advances slowly and sometimes halts transiently, indicative of disrupted or intermittent TT motion caused by elevated bending rigidity. Taken together, these observations confirm that GD associated structural remodeling leads to a spectrum of TT abnormalities, ultimately compromising the dynamic adaptability of RBCs under shear.

\begin{figure}[!tbp]
\begin{center}
\includegraphics[width=1.000\textwidth]{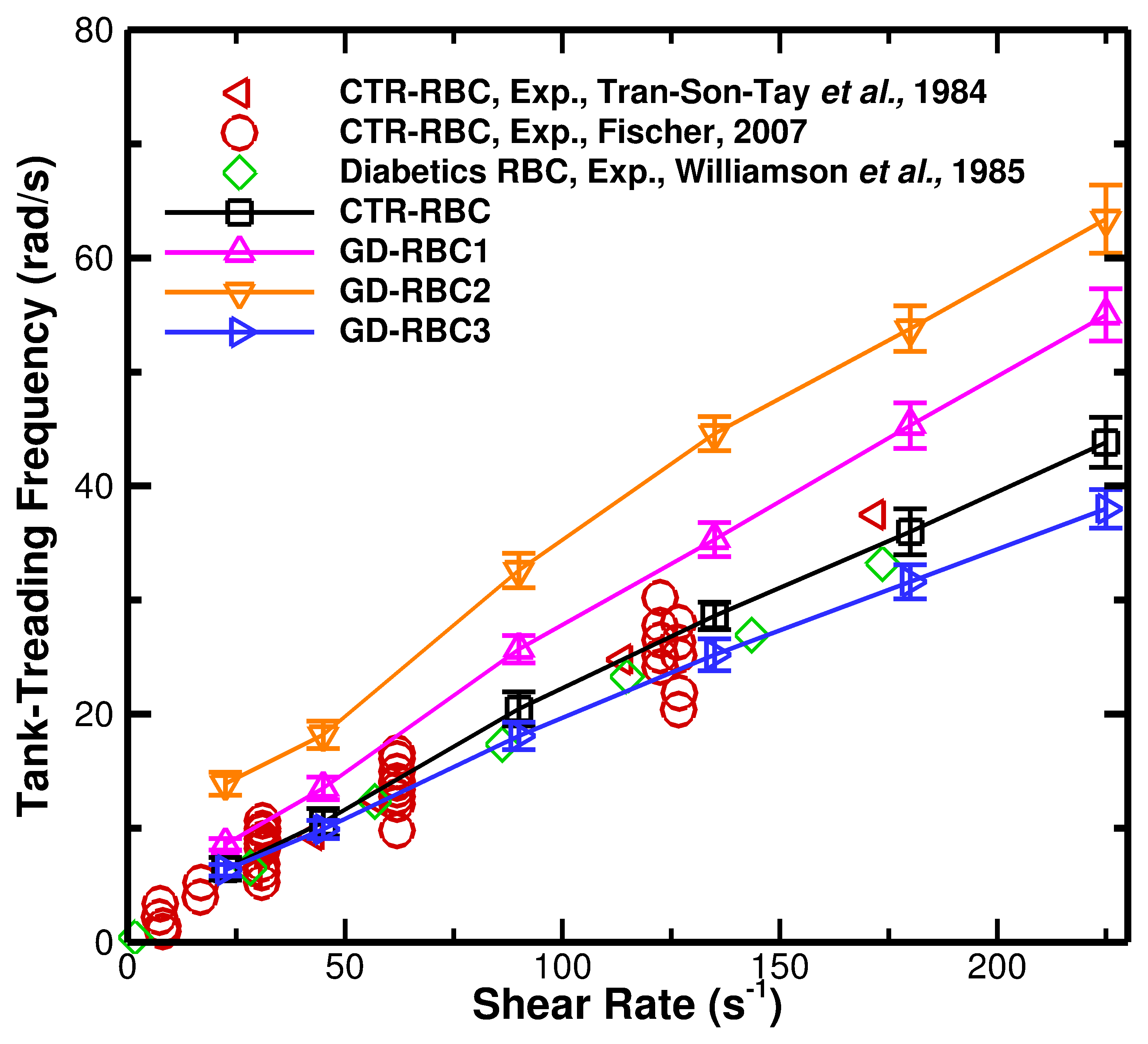}
\end{center}
\vspace{-0.15in}
\caption{\small{\bf Tank-treading frequency of CTR and GD-RBCs across shear rates.} Tank-treading frequency (rad/s) as a function of shear rate (s$^{-1}$) for CTR-RBC and GD-RBC models (GD-RBC1, GD-RBC2, GD-RBC3). Experimental data for CTR-RBCs are shown from Tran-Son-Tay et al. (1984) and Fischer (2007), with additional comparison to diabetic RBCs from Williamson et al. (1985). While frequency increases with shear rate for all cell types, GD-RBC1 and GD-RBC2 exhibit higher tank-treading frequencies than CTR-RBCs, possibly due to altered membrane tension and cell shape. GD-RBC3 shows lower frequencies across all shear rates, consistent with its elevated bending modulus, which is the only parameter distinguishing it from GD-RBC2. These findings highlight the subtype-specific mechanical responses of GD-RBCs under shear flow.}

\label{fig:tanktreading_frequency}
\end{figure}

To capture these trends quantitatively, we calculated the angular frequency of membrane rotation across a range of shear rates for each RBC type (Fig.~\ref{fig:tanktreading_frequency}). For CTR-RBCs, the tank-treading frequency increased approximately linearly with shear rate, from about 12~rad/s at 50~s$^{-1}$ to $\sim$44~rad/s at 225~s$^{-1}$, in close agreement with experimental measurements by Tran-Son-Tay et al.~\cite{tran1984determination} and Fischer~\cite{fischer2007tank}. Diabetic RBCs~\cite{williamson1985microrheologic, CHANG2017Diabete} exhibited tank-treading frequencies similar to or slightly below CTR-RBCs, falling between our CTR and GD-RBC3 cases. Malaria-infected RBCs, by contrast, are known to undergo strong slow-down or even transition into tumbling~\cite{peng2015stability}, consistent with the extreme dynamics we observe in GD-RBC3. Interestingly, GD-RBC1 and GD-RBC2 exhibited higher frequencies than controls at equivalent shear rates, indicating that altered geometry and reduced surface-to-volume ratio can promote faster membrane rotation despite increased stiffness. For example, at 100~s$^{-1}$, CTR-RBCs rotated at $\sim$22~rad/s, compared to $\sim$28~rad/s for GD-RBC1 (a $\sim$25\% increase) and $\sim$35~rad/s for GD-RBC2 (a $\sim$58\% increase). At higher shear (200~s$^{-1}$), CTR-RBCs reached $\sim$40~rad/s, whereas GD-RBC2 approached $\sim$58~rad/s. By contrast, GD-RBC3 consistently showed reduced frequencies relative to CTR-RBC ($\sim$20~rad/s vs. $\sim$22~rad/s at 100~s$^{-1}$), reflecting that its elevated bending modulus restricts membrane motion. These results suggest that changes in shape and shear modulus in GD-RBCs may enhance rotational dynamics under shear, despite increased bending modulus. These results highlight that tank-treading dynamics are jointly governed by shear modulus, bending modulus, and surface-to-volume ratio, leading to a non-monotonic relationship between GD severity and dynamic behavior. In GD, elevated bending and shear modulus, combined with shape abnormalities, lead to mechanical constraints that hinder the full execution of tank-treading cycles. The altered motion not only reflects impaired deformability but may also have functional consequences: efficient tank-treading minimizes flow disturbance and hydrodynamic drag, and its impairment may exacerbate circulatory resistance and promote splenic entrapment. 

\subsection*{RBC traversal dynamics in a capillary-like microchannel}

\begin{figure}[!tbp]
\begin{center}
\includegraphics[width=1.000\textwidth]{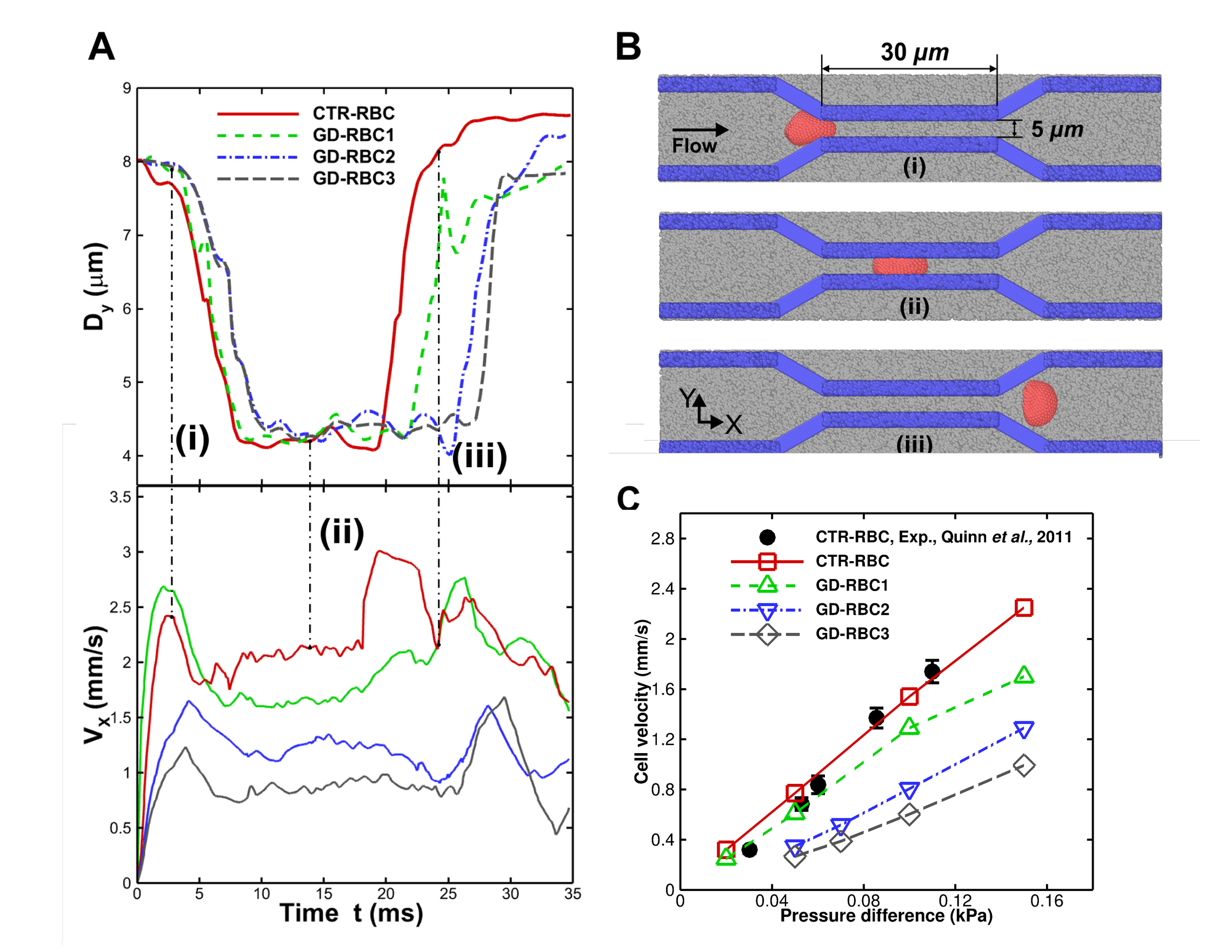}
\end{center}
\vspace{-0.15in}
\caption{\small{\bf Deformation and velocity of CTR and GD-RBCs in a microchannel constriction.} (A) Time-dependent changes in transverse diameter (\(D_y\)) and velocity (\(V_x\)) of CTR-RBC and GD-RBCs (GD-RBC1, GD-RBC2, GD-RBC3) as they pass through a constriction in the microchannel under a driving pressure difference of 0.15 kPa, with regions (i), (ii), and (iii) representing different stages of constriction traversal. (B) Simulated snapshots of RBCs at key positions (i), (ii), and (iii) within the microchannel, showing differences in shape and orientation for CTR-RBC. The channel dimension is denoted in the figure, where the depth of the channel is 2.7 $\mu$m (data not shown). Initially, all cells were located at the entrance of the channel (left end of the microfluidic channel). (C) Relationship between cell velocity (defined as the average of \(V_x\) over time) and pressure difference across the channel for CTR-RBC and GD-RBC models, with experimental data from Quinn et al.~\cite{quinn2011combined} for comparison. GD-RBCs show reduced velocities compared to CTR-RBC, indicating increased rigidity and resistance to flow under pressure.} 
\label{fig:microchannel}
\end{figure}

To examine how Gaucher disease alters RBC deformability under physiologically relevant confinement, we simulated the passage of individual cells through a narrow microfluidic constriction designed to approximate capillary-scale dimensions. The constricted segment of the channel measures $30~\mu$m in length ($x$), $5~\mu$m in width ($y$), and $2.7~\mu$m in depth ($z$), slightly smaller than the unstressed RBC diameter, requiring cells to undergo substantial deformation in order to pass through. Time-resolved measurements of cell thickness ($D_y$) and axial velocity ($V_x$) revealed striking differences between control and GD-RBCs (Fig.~\ref{fig:microchannel}A). CTR-RBCs exhibited rapid and pronounced thinning upon entry, with $D_y$ decreasing from $\sim$8.0~$\mu$m to $\sim$4.2~$\mu$m (a $\sim$47.5\% reduction), and recovered back to baseline after exiting the constriction. Their velocity remained relatively high throughout ($V_x \approx 2.0$--2.5~mm/s), reflecting smooth transit and efficient deformation. In contrast, GD-RBCs showed progressive impairment in both deformation and velocity. GD-RBC1 demonstrated a slower onset of deformation and a noticeable reduction in transit speed, with $D_y$ remaining below $4.5~\mu$m for approximately 13~ms and velocities reduced to around 1.8~mm/s, indicating difficulty adapting to the narrowing geometry. The mechanical limitations were even more pronounced in GD-RBC2 and GD-RBC3: their $D_y$ values plateaued below $4.5~\mu$m for nearly 15~ms, indicating prolonged traversal, while their velocities dropped to $\sim$1.2 and $\sim$0.9~mm/s, respectively. Notably, GD-RBC3 experienced near-stalling during constriction entry, with $D_y$ remaining below $4.5~\mu$m for approximately 17~ms and with $V_x$ falling to $\sim$0.5~mm/s---less than 40\% of the control value. Together, these findings highlight a severity-dependent decline in the ability of cells to adapt to confined flow.

To delineate the traversal dynamics of RBCs through the constriction, we examined cell morphologies at three characteristic positions: entry (i), midpoint (ii), and exit (iii), as shown by the simulation snapshots in Fig.~\ref{fig:microchannel}B. The CTR-RBC adapts smoothly to the narrowed geometry: upon entering (i), it elongates into a bullet-like profile, squeezes efficiently through the midpoint (ii), and recovers its original biconcave shape upon exiting (iii). This streamlined transition reflects high deformability and effective alignment with the flow direction. In contrast, GD-RBCs exhibit progressive impairments. GD-RBC1 is able to pass through the constriction but deforms less efficiently, adopting a blunted profile at stage (ii) and requiring nearly twice the transit time of CTR-RBC. GD-RBC2 shows more pronounced resistance: it remains wedged near the entrance for an extended period, progressing only slowly toward the midpoint, with its centroid displacement plateauing for hundreds of milliseconds before eventual passage. GD-RBC3 displays the most severe dysfunction: although it eventually traverses the constriction, it does so extremely slowly, with prolonged delays at both entry and midpoint stages and velocities falling below 0.5~mm/s—less than 25\% of the CTR-RBC value. These stage-resolved comparisons confirm that GD-related increases in membrane stiffness and reductions in surface-to-volume ratio ($S/V$) critically impair the ability of cells to undergo large shape transformations required for capillary traversal. The severity-dependent sequence—smooth passage (CTR-RBC), delayed adaptation (GD-RBC1), slow wedging passage (GD-RBC2), and near-stalled transit (GD-RBC3)—provides direct visual and quantitative evidence of how cellular heterogeneity in Gaucher disease translates into microvascular flow resistance.  

To further quantify flow resistance, we computed the relationship between applied pressure drop and resulting RBC velocity across the channel (Fig.~\ref{fig:microchannel}C). CTR-RBCs displayed a near-linear pressure–velocity relationship, consistent with deformable bodies whose flow resistance scales smoothly with applied force, reaching $\sim$2.3~mm/s at $\Delta P = 0.15$~kPa, in agreement with experimental measurements by Quinn et al.~\cite{quinn2011combined}. In contrast, GD-RBC1 plateaued at only $\sim$1.7~mm/s under the same pressure (a $\sim$24\% reduction), GD-RBC2 reached $\sim$1.3~mm/s (a $\sim$43\% reduction), and GD-RBC3 fell to $\sim$1.0~mm/s (a $\sim$56\% reduction). Notably, GD-RBC2 and GD-RBC3 displayed threshold-like behavior, showing little to no motion below $\Delta P = 0.05$~kPa and only sluggish transit even at higher pressures, whereas GD-RBC1 retained an approximately linear pressure--velocity response. Such nonlinear responses reflect extreme mechanical resistance and limited capacity for shape adaptation.

\subsection*{RBC traversal through splenic slits}

The spleen imposes one of the most stringent mechanical filters on circulating RBCs, forcing them to deform through narrow interendothelial slits (IES) in the splenic red pulp~\cite{macdonald1987kinetics,safeukui2018sensing}. These slits, typically 1–3~$\mu$m in width, act as biomechanical checkpoints where only sufficiently flexible cells can pass; those with reduced deformability are typically retained and phagocytosed \cite{safeukui2012quantitative}. This mechanical sensing has been implicated in the development of splenomegaly and other hematological complications in diseases such as Gaucher and hereditary spherocytosis~\cite{pivkin2016biomechanics,Li2018Mechanics}. To probe how Gaucher disease impairs RBC deformability under such physiological constraints, we implemented a numerical model of cell traversal through an IES with physiologically representative geometry: a slit height of 1.2~$\mu$m, width of 5.0~$\mu$m, and depth of 2.5~$\mu$m. Each RBC was initialized 9~$\mu$m upstream of the slit entrance, and a constant pressure gradient of 6~Pa$\cdot\mu$m$^{-1}$ was applied to drive flow in the $x$-direction—within the physiological range of intra-splenic pressures \cite{atkinson1954intrasplenic,pivkin2016biomechanics}. This setup recapitulates the spleen’s stringent deformability screening, where only sufficiently flexible cells can transit slits as narrow as $\sim$2--3~$\mu$m, while more rigid cells are sequestered and cleared by splenic macrophages~\cite{safeukui2018sensing}.

\begin{figure}[!tbp]
\begin{center}
\includegraphics[width=1.000\textwidth]{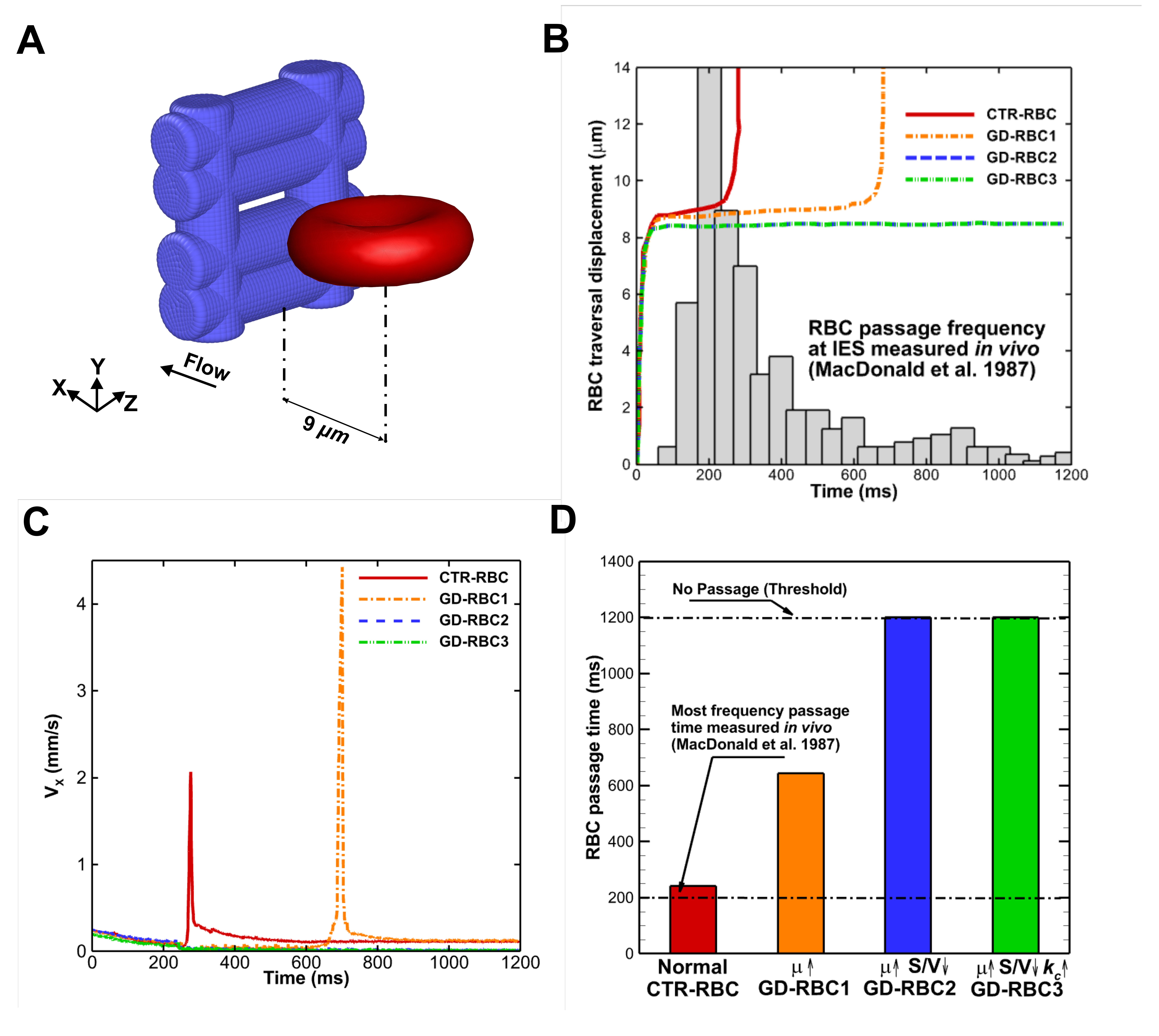}
\end{center}
\vspace{-0.15in}
\caption{\small{\bf Passage of CTR and GD-RBCs through an interendothelial slit (IES) model.} 
(A) Simulation setup showing an RBC approaching the IES. Flow is directed along the X-axis. (B) RBC traversal displacement over time for CTR-RBC and GD-RBCs (GD-RBC1, GD-RBC2, GD-RBC3), with in vivo transit time data from MacDonald et al. (1987) for comparison. GD-RBCs exhibit delayed or obstructed passage relative to CTR-RBC due to increased rigidity and reduced deformability. (C) RBC velocity in the X-direction ($V_x$) over time. CTR-RBC moves smoothly through the IES, whereas GD-RBCs show slower velocities, with GD-RBCs showing prolonged stagnation due to altered mechanical properties. (D) RBC passage time comparison. CTR-RBC completes passage within the physiological window observed in vivo, GD-RBC1 is delayed but completes, whereas GD-RBC2 and GD-RBC3, characterized by reduced surface-area-to-volume ratio and increased rigidity, exceed the passage threshold, indicating a high likelihood of IES blockage.}

\label{fig:splenic_slit}
\end{figure}

Fig.~\ref{fig:splenic_slit}B shows traversal displacement over time. CTR-RBCs completed passage in $\sim$250$\,\pm$50~ms, closely matching in vivo IES transit times ($\sim$200~ms) reported by MacDonald et al.~\cite{macdonald1987kinetics}. In contrast, GD-RBCs demonstrate progressively impaired filtration. GD-RBC1, the least affected subtype, required $\sim$600--700~ms to traverse—more than twice the duration of CTR-RBCs. GD-RBC2 and GD-RBC3 were the most impaired, with traversal times exceeding 1200~ms and stalling behavior at the slit entrance. The traversal dynamics are further illustrated in Fig.~\ref{fig:splenic_slit}C. CTR-RBC accelerated rapidly to $V_x \approx 2.1$~mm/s at $\sim$270~ms, consistent with Fig.~\ref{fig:splenic_slit}B. GD-RBC1 shows initial deceleration due to resistance to deformation. GD-RBC1 peaked to $V_x \approx 4.4$~mm/s at $\sim$600--700~ms. GD-RBC2 and GD-RBC3 exhibit stagnation. These differences arise from reductions in surface-to-volume ratio and increases in bending modulus that prevent GD-RBCs from assuming the tubular configurations necessary for narrow-slit passage. Comparison of total passage times (Fig.~\ref{fig:splenic_slit}D) highlights this severity-dependent hierarchy. CTR-RBCs fell within the physiological passage window ($\sim$250~ms), GD-RBC1 (shear modulus increase) was delayed to $\sim$600--700~ms, whereas GD-RBC2 (shear modulus increase and reduced $S/V$ ratio) and GD-RBC3 (shear modulus increase, reduced $S/V$ ratio, and elevated bending modulus) exceeded 1200~ms, effectively reaching the no-passage threshold and suggesting sequestration by the spleen. These findings align with clinical observations of hypersplenism and splenomegaly in GD patients~\cite{franco2013abnormal,dupuis2020effects} and offer a mechanistic explanation rooted in biomechanical selection. Taken together, these results demonstrate a progressive hierarchy of deformability loss: CTR-RBC $>$ GD-RBC1 $>$ GD-RBC2/GD-RBC3.

\subsection*{Blood viscosity in GD and comparison to other hematologic disorders}

\begin{figure}[!tbp]
\begin{center}
\includegraphics[width=1.000\textwidth]{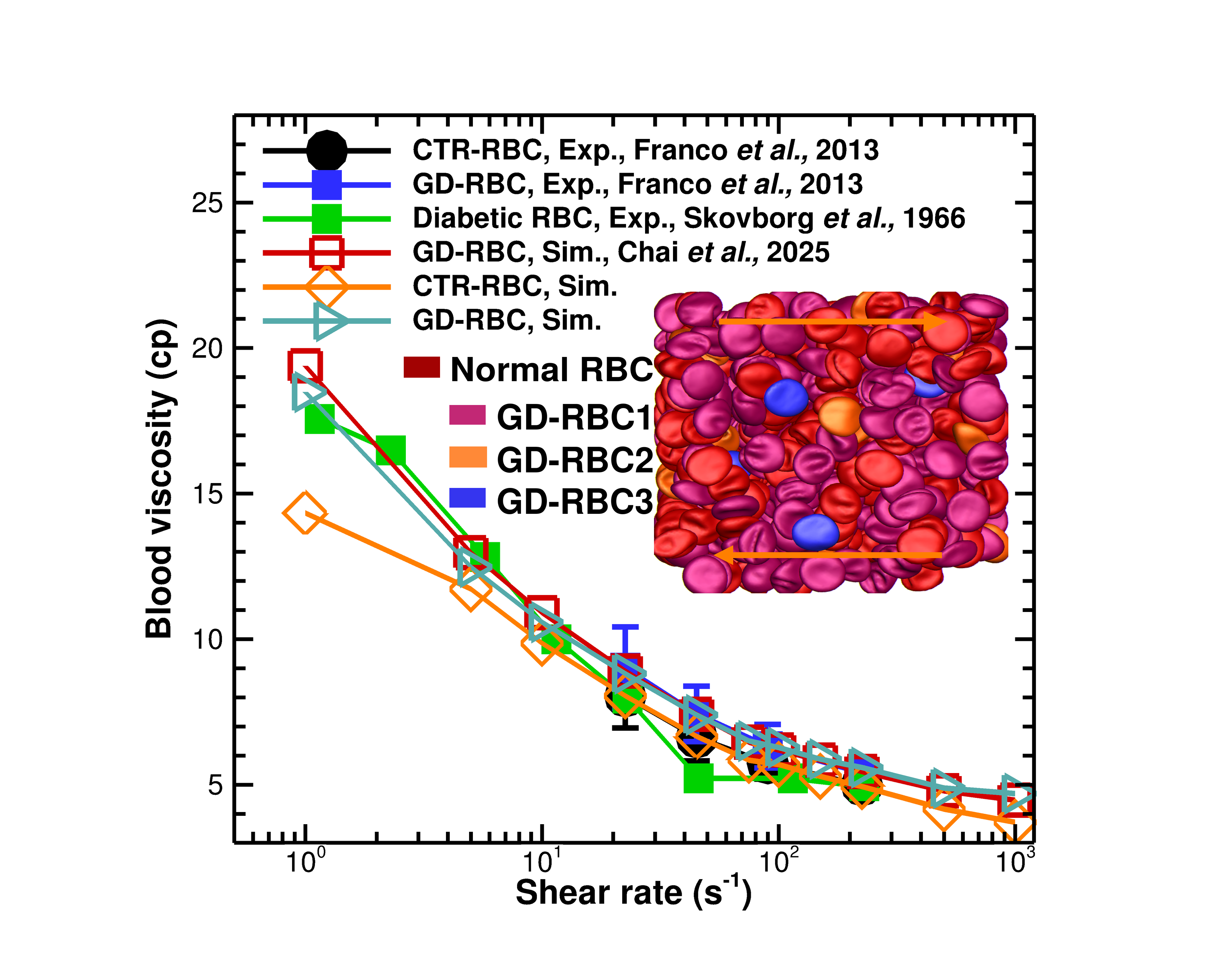}
\end{center}
\vspace{-0.15in}
\caption{\small{\bf  Blood relative viscosity of control and 
GD-RBC suspensions across shear rates, with comparison to other hematologic disorders.} Simulated blood viscosity as a function of shear rate ($\mathrm{s}^{-1}$) for CTR-RBCs and GD-RBC models (GD-RBC1, GD-RBC2, GD-RBC3), evaluated under physiologically relevant shear rates from 1 to 1000~$\mathrm{s}^{-1}$. Experimental data for CTR-RBC and GD-RBC suspensions from Franco et al.~\cite{franco2013abnormal} and for diabetic RBCs from Skovborg et al.~\cite{SKOVBORG1966129} are included for comparison. Simulation data for GD-RBC pure suspensions from our previous study~\cite{chai2025silico} are also shown. Simulations (GD-RBC, Sim.) incorporate RBC subtype proportions reported by Franco et al.~\cite{franco2013abnormal}. The good agreement between simulation and experimental viscosity, despite the low fractions of GD-RBC2 and GD-RBC3, highlights the significant contribution of these highly rigid subpopulations and validates the mixed-population modeling approach. 
\textbf{Inset:} Visualization of the RBC suspension flow field with colors indicating RBC subtypes in the proportions.} 
\label{fig:viscosity_comparison}
\end{figure}

Compared to previous modeling studies~\cite{chai2025silico}, which assumed a homogeneous RBC population, the present mixed-population model incorporates the experimentally determined proportions of GD-RBC subtypes reported by Franco et al.~\cite{franco2013abnormal}: GD-RBC with normal shape accounted for 96.0\%, GD-RBC with abnormal shape for 4.0\%. The inset in Fig.~\ref{fig:viscosity_comparison} illustrates the simulated RBC suspension flow field at $H_t=36\%$, color-coded by subtype. This mixed-population model better reflects the composition of patient blood and enables a more physiologically relevant assessment of blood rheology.

We then evaluated the relative viscosity of blood suspensions containing control and GD-RBCs across physiologically relevant shear rates (1–1000~s$^{-1}$). Fig.~\ref{fig:viscosity_comparison} shows the simulated viscosity curves for CTR-RBC and GD-RBC models (GD-RBC1, GD-RBC2, GD-RBC3), alongside experimental measurements from Franco et al.~\cite{franco2013abnormal} for GD patient blood and from Skovborg et al.~\cite{SKOVBORG1966129} for diabetic RBCs. Enhanced adhesion was modeled via a Morse potential tuned to reproduce the elevated disaggregation threshold (250 s$^{-1}$ for GD-RBC versus 110 s$^{-1}$ for control RBC) reported experimentally~\cite{franco2013abnormal, chai2025silico}. CTR-RBC suspensions exhibited the expected shear-thinning behavior of healthy blood, with viscosity decreasing from $\sim$14~cp at 1~s$^{-1}$ to $\sim$4~cp at 1000~s$^{-1}$. In contrast, GD-RBC suspensions showed consistently elevated viscosity across all shear rates, most prominently at low shear ($\lesssim 10~\mathrm{s}^{-1}$), where values reached $\sim$18.5~cp compared to $\sim$14.3~cp for controls---a $\sim$29\% increase. This pronounced low-shear elevation reflects impaired deformability and enhanced intercellular interactions among GD-RBCs. At intermediate shear rates ($\sim$10--100~s$^{-1}$), the gap between GD and control suspensions narrowed, with GD-RBC viscosity at $\sim$11--7~cp compared to $\sim$10--6~cp for controls, but remained systematically higher. At high shear ($\geq 100~\mathrm{s}^{-1}$), both groups fell toward $\sim$4--6~cp, with GD suspensions retaining a residual elevation of $\sim$10\% at 100~s$^{-1}$ that widened again to $\sim$27\% at 1000~s$^{-1}$. The trends closely parallel experimental observations in GD patient blood~\cite{franco2013abnormal} and resemble diabetic RBC suspensions~\cite{SKOVBORG1966129}, which also display a 30--40\% viscosity increase at low shear relative to controls. Such parallels emphasize cell--cell adhesion and reduced deformability as shared mechanisms impairing microvascular perfusion in both GD and diabetes mellitus~\cite{chai2025silico}.  

Notably, despite the low fractions of GD-RBC2 and GD-RBC3 in the mixed-population model introduced above, the simulated viscosity trends show good agreement with experimental measurements~\cite{franco2013abnormal}, demonstrating that these highly rigid subpopulations, although numerically minor, contribute disproportionately to viscosity elevation. 

\section*{Discussion and Summary}

Gaucher disease (GD) is associated with altered red blood cell (RBC) biophysical properties that contribute to hematologic complications, including delayed microcirculatory transit and vascular occlusions. However, the individual roles of these GD-associated RBC alterations remain unclear.
Due to the limitations of in vitro experimental measurements discussed above, we employ computational approaches. Advances in computational power over the past two decades have enabled the development of multiscale biophysical models that provide mechanistic insight into RBC behavior across length scales~\cite{Rajabi2018Experimental,tang2017openrbc,zhang2017multiscale,chai2023dynamics,chai2022periodic,chang2016md}. In this work, we established a validated DPD modeling framework to dissect how GD alters RBC mechanics and hemorheology. By systematically varying shear modulus, surface-to-volume ratio ($S/V$), and bending modulus, and by incorporating enhanced cell–cell adhesion, the model reproduced the mechanical and rheological signatures observed experimentally in GD patients. The good agreement between simulations and experimental measurements validates the approach and provides mechanistic insight into the contribution of RBC abnormalities to GD pathophysiology.

A key feature of our study is the explicit parameterization of three representative GD-RBC subtypes (GD-RBC1, GD-RBC2, GD-RBC3), each designed to reflect different aspects of biophysical properties reported in GD blood~\cite{franco2013abnormal,dupuis2020effects}. GD-RBC1 incorporates an approximately one order-of-magnitude increase in shear modulus ($9E_{s0}$ versus $E_{s0}$ for controls), consistent with the reduced elongation observed by adhesion assay experiments and our prior simulations~\cite{franco2013abnormal,chai2025silico}. This variant isolates the effect of membrane stiffening while maintaining a normal $S/V$ ratio and bending rigidity. In conjunction with the morphological changes reported earlier, such an increase in shear modulus is expected to hinder GD-RBC transit through narrow splenic slits and capillaries, leading to microcirculatory impairment, splenomegaly, and associated vascular complications in Gaucher disease~\cite{Stirnemann2017Pathophysiology}. GD-RBC2 combines increased shear modulus with a reduced $S/V$ ratio (1.22 versus 1.44), reflecting the morphological evidence of spheroidal shapes and schistocyte-like cells in GD patients. These morpho-biophysical alterations may impair circulation and promote splenic retention, and our simulation results below support this hypothesis, in line with reported clinical complications such as splenomegaly and ischemic bone events~\cite{franco2013abnormal,dupuis2020effects}. GD-RBC3 incorporates all three alterations: increased shear modulus, reduced $S/V$, and a doubled bending modulus ($k_c = 4.8\times10^{-19}$ J versus $2.4\times10^{-19}$ J). This configuration represents the most severe phenotype, motivated by lipidomic data showing sphingolipid enrichment that stiffens curvature elasticity~\cite{dupuis2020effects}, thereby limiting large-scale bending deformations. Together, these subtypes allow us to isolate the contributions of stiffness, geometry, and curvature resistance, and to evaluate their combined impact on RBC behavior in physiologically relevant flow conditions. More generally, the reduction of $S/V$ ratio reflects the effect of surface area loss, which is known to impair RBC passage through narrow constrictions, as quantitatively demonstrated in closely related settings for stomatocytes~\cite{chai2026stomatocyte} and for mechanically heterogeneous reticulocyte populations in confined microchannel flow~\cite{chai2026reticulocyte}.

Our results show that GD-RBCs exhibit marked morphological alterations, with reduced projected surface area and lower $S/V$ ratio compared to controls. These changes were more pronounced in GD-RBC2 and GD-RBC3, directly linking geometry to reduced flexibility. Our model reproduces the dependence of cell shape on the S/V ratio, capturing the transition from biconcave to increasingly spheroidal morphologies. Ektacytometry confirmed impaired elongation indices at low shear stresses ($<3$ Pa), consistent with the simulation findings that stiffened membranes fail to deform under weak shear stresses. At higher stresses, differences diminished as cytoplasmic viscosity dominated deformation, in line with prior guidelines~\cite{baskurt2009guidelines}. Optical tweezers stretching further highlighted severity-dependent reductions in axial elongation: GD-RBC1 showed moderate impairment, while GD-RBC3 displayed a 27\% reduction in axial diameter at 100 pN. These findings reproduce the spectrum of RBC deformability reported in GD patients. Simulations of membrane dynamics under shear revealed subtype-specific effects. GD-RBC3 exhibited delayed and irregular tank-treading, attributable to its elevated bending modulus. In contrast, GD-RBC1 and GD-RBC2 displayed unexpectedly higher tank-treading frequencies compared to controls. This counterintuitive effect arises because reduced $S/V$ and increased shear modulus together promote more compact cell shapes that rotate faster in shear flow. Our results extend earlier reports of tank-treading slow-down in diseased RBCs, such as those affected by diabetes~\cite{CHANG2017Diabete}, and provide quantitative benchmarks for how GD severity correlates with the loss of dynamic membrane adaptability. This mechanistic insight underscores the relevance of tank-treading frequency as a biomarker of RBC mechanical health in Gaucher disease~\cite{matteoli2021impact,dupire2012full}. Our results also emphasize that GD-induced alterations interact nonlinearly: geometry changes can mask or even reverse the expected impact of stiffening on dynamic behavior. Such insights highlight the importance of integrating multiple parameters when interpreting GD-RBC mechanics. 

At the single GD-RBC level, microchannel constriction revealed severity-dependent impairments consistent with experimental microfluidic observations. In the microchannel constriction model, GD-RBC2 and GD-RBC3 showed prolonged residence times, reduced velocities, and in some cases near-stalling, while GD-RBC1 exhibited only moderate delay. These simulation results are consistent with prior microfluidic experiments, which report that pathologically stiffened cells—such as those affected by heat treatment or sickling—display reduced transit velocities and higher channel occlusion rates~\cite{quinn2011combined, baskurt2003blood}. Interendothelial slit simulations further underscored the vulnerability of GD-RBCs to mechanical filtration. CTR-RBCs traversed slits within approximately 250 ms, consistent with in vivo reports~\cite{macdonald1987kinetics}. GD-RBC1 required more than twice that time, whereas GD-RBC2 and GD-RBC3 exceeded 1200 ms, effectively failing to pass. Even though GD-RBC2 and GD-RBC3 comprise only $\sim$4.0\% of the total population, their extreme rigidity and prolonged transit disproportionately impair splenic filtration, consistent with the spleen’s role as both a sensor and effector in Gaucher disease pathophysiology \cite{safeukui2018sensing,deplaine2011sensing,asaro2021red}. Such mechanical filtration deficits are also suggested to couple to macrophage-mediated RBC clearance, an interplay supported by Dupuis et al.~\cite{dupuis2022phagocytosis}, who demonstrated enhanced in vitro phagocytosis of GD-RBCs by macrophages, and that we examine within an integrated multiscale signaling--biophysical framework spanning sickle cell and Gaucher disease~\cite{chai2026multiscale}. A forthcoming companion study combining the present simulation framework with experimental microsphiltration assays is expected to further substantiate this splenic-retention link. These multiscale impairments provide a mechanistic link between altered RBC biomechanics and clinical manifestations in GD. Impaired microcirculatory transit and increased flow resistance promote vascular occlusions, with splenic retention of rigid RBCs driving hypersplenism and splenomegaly.

At the suspension level, mixed-population simulations recapitulate the elevated blood viscosity observed in GD patient samples and show good agreement with experimental measurements when incorporating the observed subtype distribution. Importantly, even though GD-RBC with abnormal shape collectively represented only about 4\% of the population, their extreme rigidity disproportionately elevated viscosity. This aligns with prior evidence that a small fraction (1\% - 5\%) of rigidified spheres can strongly increase flow resistance~\cite{perazzo2022effect,kuck2022impact}. This finding is particularly relevant to GD, where only a subset of RBCs may exhibit severe rigidity (GD-RBC2/3), yet their mechanical dominance could impair bulk flow and increase perfusion heterogeneity. Comparisons with diabetic and malaria-infected RBCs contextualized GD within the broader spectrum of hemorheological disorders: aggregation-driven viscosity resembles diabetes, while severe rigidity resembles malaria-infected cells~\cite{baskurt2003blood, nader2019blood, cooke2001malaria}. In a closely related framework, our recent companion analysis of dehydrated stomatocytes (xerocytosis) likewise reveals geometry-driven viscosity elevation at low shear quantitatively comparable to GD hyperviscosity~\cite{chai2026stomatocyte}, reinforcing the view that mechanically distinct RBC phenotypes can converge on similar bulk hemorheological signatures. An important caveat is that the morphological heterogeneity captured by our subtype models is directly supported by imaging of GD patient blood (Fig.~\ref{fig:morphology}), whereas the underlying heterogeneity in membrane lipid burden is inferred indirectly, from age-dependent accumulation of sphingolipids during the RBC lifespan~\cite{dupuis2020effects}. With that distinction in mind, these findings emphasize the importance of accurately capturing RBC heterogeneity when interpreting hemorheological abnormalities in GD, and highlight the quantitative link between specific mechanical impairments and flow dysfunction.

Experimental measurements often confound co-varying factors and make it difficult to isolate contributions of geometry, shear modulus, and bending modulus. In contrast, our simulations enable systematic, independent perturbations of these parameters to disentangle their effects on RBC mechanics and flow behavior. This in silico strategy clarifies several otherwise entangled observations. First, reductions in $S/V$ alone reproduce the loss of biconcavity and prolong capillary traversal by increasing cell thickness, even without additional membrane stiffening. Second, increasing shear modulus at fixed geometry can transiently accelerate tank-treading (TT) frequency by reducing elongation amplitude, but also destabilizes periodicity at higher values, explaining the non-monotonic TT responses observed across GD severities. Third, elevated bending rigidity predominantly slows and intermittently disrupts TT motion, while prolonging passage times through splenic slits beyond the functional threshold, thereby identifying bending resistance as a principal constraint on dynamic adaptability under shear. At the suspension scale, the combined effects of these parameters reflect heterogeneous mixtures, where a small rigid subpopulation disproportionately increases bulk flow resistance. Together, these orthogonal perturbations establish a mechanistic map linking parameter space to phenotypic outcomes. By integrating single-cell measures with cell bulk properties within the same parametric framework, the simulations offer a unifying perspective for interpreting heterogeneous clinical observations.

Our model assumes uniform parameters for each GD subtype and does not explicitly incorporate additional blood components, such as leukocytes, platelets, or the detailed architecture of splenic vasculature. Despite these simplifications, the study establishes a validated multiscale computational framework. Future work will integrate patient-specific RBC properties and realistic vascular geometries to further improve predictive capability. Nonetheless, by systematically varying geometry, shear modulus, and bending rigidity in isolation, the model disentangles co-varying effects that are difficult to resolve experimentally, providing mechanistic links between single-cell morphology and dynamics and population-level viscosity. Moreover, this methodological framework is broadly generalizable to other hemolytic and storage disorders involving altered RBC mechanics, offering a versatile platform for mechanistic insight and the development of patient-specific therapeutic strategies.

\section*{Supporting Material}

An online supplement to this article can be found by visiting BJ Online at \url{http://www.biophysj.org}.

\section*{Author Contributions}

Z.C., M.F., P.A.B., and G.E.K. conceived and designed the research. Z.C. developed the computational framework and performed all simulations. M.d.P. and M.F. conducted the experiments. Z.C., M.F., and G.E.K. analyzed the data and interpreted the results. M.F., P.A.B., and G.E.K. supervised the project. All authors contributed to writing and revising the manuscript and approved the final version.

\section*{Acknowledgments}
We acknowledge support from the National Institutes of Health (Grant No. R01HL154150) and from the France 2030 program through the Idex Universit\'e Paris Cit\'e (ANR-18-IDEX-0001, GR-Ex). We thank the Centre de R\'ef\'erence des Maladies Lysosomales (CRML, H\^opital Beaujon, AP-HP) for clinical support and for providing access to the Gaucher patient cohort that underpins the lipidomic and deformability data used in this work. Simulations were carried out at the Center for Computation and Visualization of Brown University.
\section*{Declaration of Interests}
The authors declare no competing interests.

\section*{Data Availability}
All data supporting the findings of this study are contained within the main text and the Supporting Material.


\end{document}